\documentclass[journal]{IEEEtran}
\usepackage{amsmath,amsfonts}
\usepackage{algorithmic}
\usepackage{algorithm}
\usepackage{array}
\usepackage[caption=false,font=normalsize,labelfont=sf,textfont=sf]{subfig}
\usepackage{textcomp}
\usepackage{stfloats}
\usepackage{url}
\usepackage{verbatim}
\usepackage{graphicx}
\usepackage{cite}
\usepackage{multirow}
\usepackage{booktabs}
\usepackage{hyperref}

\begin{document}

\title{PAGNet: Pluggable Adaptive Generative Networks for Information Completion \\ in Multi-Agent Communication}

\author{Zhuohui Zhang,~\IEEEmembership{Graduate Student Member,~IEEE,} Bin Cheng, \IEEEmembership{Member,~IEEE,} Zhipeng Wang, \IEEEmembership{Member,~IEEE,} Yanmin Zhou, \IEEEmembership{Member,~IEEE,} Gang Li, \IEEEmembership{Member,~IEEE,} Ping Lu, \\Bin He, \IEEEmembership{Senior Member,~IEEE,} Jie Chen,~\IEEEmembership{Fellow,~IEEE}
\thanks{This work was supported by the National Natural Science Foundation of China under grants 62103302 and U1713215, by the Shanghai Chenguang Program under grant 22CGA19, and by the Shanghai Municipal Science and Technology Major Project under grant 2021SHZDZX0100. (Corresponding author: Bin Cheng.)}

\thanks{The authors are with the Department of Control Science \& Engineering, Tongji University, Shanghai 201804, China, and also with the National Key Laboratory of Autonomous Intelligent Unmanned Systems, Shanghai 201203, China (e-mail: \{zhangzhuohui; bincheng; wangzhipeng; yanmin.zhou; lig; pinglu; hebin; chenjie206\}@tongji.edu.cn).}}

\markboth{Journal of \LaTeX\ Class Files}%
{Shell \MakeLowercase{\textit{et al.}}: A Sample Article Using IEEEtran.cls for IEEE Journals}

\IEEEpubid{0000--0000/00\$00.00~\copyright~2021 IEEE}

\maketitle

\begin{abstract}
For partially observable cooperative tasks, multi-agent systems must develop effective communication and understand the interplay among agents in order to achieve cooperative goals. However, existing multi-agent reinforcement learning (MARL) with communication methods lack evaluation metrics for information weights and information-level communication modeling. This causes agents to neglect the aggregation of multiple messages, thereby significantly reducing policy learning efficiency. In this paper, we propose pluggable adaptive generative networks (PAGNet), a novel framework that integrates generative models into MARL to enhance communication and decision-making. PAGNet enables agents to synthesize global states representations from weighted local observations and use these representations alongside learned communication weights for coordinated decision-making. This pluggable approach reduces the computational demands typically associated with the joint training of communication and policy networks. Extensive experimental evaluations across diverse benchmarks and communication scenarios demonstrate the significant performance improvements achieved by PAGNet. Furthermore, we analyze the emergent communication patterns and the quality of generated global states, providing insights into operational mechanisms.

\end{abstract}

\begin{IEEEkeywords}
Multi-agent reinforcement learning, Multi-agent communication, Generative adversarial network, Pluggable architecture
\end{IEEEkeywords}

\section{Introduction}

\IEEEPARstart{T}{he} field of multi-agent reinforcement learning (MARL) experiences a recent surge in interest due to its potential for modeling and solving complex cooperative tasks in real-world applications such as autonomous vehicles, robot swarms, smart grid control, and distributed sensor networks \cite{zhou2021smarts, huttenrauch2019deep, roesch2020smart, hu2023mo}. 
Unlike single-agent reinforcement learning, MARL faces the challenge of promoting effective coordination among agents to achieve a common objective \cite{eccles2019biases, lin2021learning, yu2022surprising}. A natural approach to cooperative MARL regards the multi-agent system as a whole using single-agent learning techniques \cite{liu2024interaction}. However, centralized methods encounter scalability challenges and limitations associated with centralized controllers \cite{foerster2018counterfactual}. Another approach is decentralized, where each agent learns its policy with single-agent techniques, addressing the limitations of centralized controllers but introducing non-stationarity and credit assignment problems \cite{lowe2017multi}. This added dimension of coordination presents significant challenges, as neither centralized \cite{foerster2018counterfactual, wang2021towards} nor decentralized \cite{li2022metadrive} methods can effectively handle both the non-stationarity caused by concurrent learning of multiple policies and scalability with an increasing number of agents. Existing MARL algorithms rely mainly on centralized training with decentralized execution (CTDE) \cite{rashid2020monotonic, sunehag2017value, lowe2017multi, lyu2021contrasting}. However, this paradigm is not without limitations. In centralized control, the assumption of access to global states is idealized \cite{wang2019learning, wang2020qplex}, while in decentralized execution, agents are limited to making decisions based on their own local observations, which may hinder their ability to make informed decisions and coordinate effectively \cite{ding2020learning, yuan2022multi}.

\IEEEpubidadjcol
To address this issue, inspired by human cooperation, a growing body of research has focused on integrating communication mechanisms within MARL frameworks \cite{foerster2016learning, sukhbaatar2016learning}. Effective communication underpins cooperation. By enabling agents to exchange information regarding their local observations, beliefs, or intentions, communication can facilitate the inference of global states and promote coordinated decision-making. Nonetheless, existing methods face three major challenges. First, the construction of communication models often lacks precision, as these models typically predefine the mode of communication, either through broadcasting \cite{sukhbaatar2016learning, jiang2018learning} or peer-to-peer mechanisms \cite{singh2018learning, jiang2018graph, wang2019learning, ding2020learning}. Although these methods address when to communicate and clarify communication targets, they fail to increase communication to the information level. Second, many methods suffer from computational inefficiencies \cite{niu2021multi}. The communication model and policy network can only jointly learn through sparse reinforcement learning rewards, which fails to separate the communication module from the training process, resulting in decreased training efficiency. Finally, communication is often limited to the decentralized execution phase \cite{liu2020multi, guan2022efficient}. During the centralized training process, these methods still rely on access to global states, which is challenging and requires significant skill to define, failing to bridge the gap between training and execution through communication.

Our motivation stems from the realization that if agent's local observations play a crucial role in generating global states information, then the corresponding communication information from that agent is also important for other agents' decision-making processes. In this paper, we propose a novel network named the pluggable adaptive generative network (PAGNet), which integrates generative networks \cite{ho2020denoising} with communication in MARL. PAGNet allows agents to generate global states from weighted local observations and utilize these states along with learned communication weights to make coordinated decisions. The term ``pluggable'' implies that PAGNet can be pretrained and seamlessly integrated into various MARL algorithms. ``Adaptive'' refers to PAGNet's ability to learn effective communication information-level weights tailored to specific MARL tasks, while ``generative'' signifies the use of generative models to synthesize global states information. During centralized training, the generative model can adaptively select weights for different agents’ local observations to better generate the global states, thus reducing the need for access to true global states. In the decentralized execution process, the policy network can access the weights in the generative model to achieve information-level weighted extraction, improving the efficiency of policy learning. Furthermore, the pluggable nature of PAGNet enables the algorithm to seamlessly transition between online and offline learning \cite{lyu2022mildly}. To evaluate the efficacy of our proposed method, we perform a comprehensive empirical evaluation on a diverse array of cooperative multi-agent benchmarks. Specifically, we assess our method on the level-based foraging (LBF) environment \cite{rangwala2020learning}, the Hallway scenario \cite{wang2019learning}, and eight distinct maps from the StarCraft multi-agent challenge (SMAC) \cite{samvelyan2019starcraft}. The experimental findings demonstrate that our PAGNet achieves superior performance compared to prior state-of-the-art methods, strong baselines, and ablated variants of our technique. Additionally, we analyze the emergent communication patterns and the quality of the generated global information, shedding light on the interpretability and robustness of our proposed networks.

The main contributions of this paper are listed as follows.

\begin{enumerate}

\item We propose a novel network that seamlessly integrates generative models within MARL for communication. This allows agents to generate global states from weighted local observations, enabling them to use these states along with learned communication weights to make coordinated decisions.

\item The adaptive information-level modeling of communication achieves an effective weighing and integration of communication information from different agents, tailoring it to specific multi-agent reinforcement learning tasks with communication, thereby bridging the gap between training and execution through seamless information exchange.

\item The pluggable design of PAGNet enables pretraining and smooth integration into different MARL algorithms involving communication. By incorporating generative model loss functions, the training of the communication module is decoupled from sparse reinforcement learning rewards, thereby enhancing the efficiency of the policy network’s learning process.

\item We present a comprehensive set of experimental evaluations across various benchmarks and communication scenarios, which substantiate the significant performance gains achieved by our proposed methodology. Additionally, visualizing the adaptive weights of communication and generating global states sheds light on the operational mechanisms of PAGNet.

\end{enumerate}

The outline of this paper is as follows: In Sec. \ref{S_2} and \ref{S_3}, we introduce the related work and preliminaries. Sec. \ref{S_4} details the proposed PAGNet framework, including the information-level weight network, the adaptive generative network, and the overall pluggable architecture design. In Sec. \ref{S_5}, we present comprehensive experimental evaluations across diverse benchmarks, analyzing PAGNet's performance, emergent communication patterns, global state generation quality, and scalability aspects. Finally, Sec. \ref{S_6} concludes the paper and discusses future work.

\section{Related Works}
\label{S_2}

Multi-agent systems are widespread in life and lie at the intersection of game theory and artificial intelligence in general, which receive much attention from researchers. Communication is essential for MARL to capture inter-agent action dependencies and is proven to enhance exploration and team rewards \cite{apicella2012social, li2021structured, zhang2020succinct}. Agents need to enhance action coordination by learning to communicate with other agents and process the message representations they receive. To address the distributed control and non-stationarity challenges in MARL, numerous methods have achieved notable progress in recent years. Our work builds upon prior research on MARL and communication mechanisms.

\subsection{MARL without Communication}

Existing MARL approaches can be divided into two categories based on the presence or absence of a communication mechanism. MARL without communication utilizes the CTDE paradigm to tackle multi-agent cooperation problems. This approach has been successfully implemented with both policy-based and value-based algorithms. CTDE allows information sharing during training while ensuring that policies are conditioned solely on the agents' local observations, thus enabling decentralized execution. Various CTDE algorithms have been developed, each offering unique advantages and addressing specific challenges in MARL.

Policy-based MARL methods include centralized policy gradient algorithms where each agent comprises a decentralized actor and a centralized critic. A notable example is multi-agent deep deterministic policy gradient (MADDPG) \cite{lowe2017multi}. MADDPG extends the deep deterministic policy gradient (DDPG) \cite{silver2014deterministic} algorithm for MARL by conditioning the actor on the history of local observations and training the critic on joint observations and actions to approximate the joint state-action value function. Another significant policy-based method is the counterfactual multi-agent (COMA) \cite{foerster2018counterfactual} policy gradient proposed, which modifies the advantage function in the actor’s loss computation to perform counterfactual reasoning for credit assignment in cooperative MARL. Multi-agent advantage actor-critic (MAA2C) and multi-agent proximal policy optimization (MAPPO) \cite{yu2022surprising} are additional examples, with MAPPO offering enhanced learning efficiency by performing several update epochs pertraining batch.

Value-based MARL methods focus on decomposing the joint state-action value function into individual state-action value functions, adhering to the individual-global-max (IGM) principle. Value decomposition networks (VDN) \cite{sunehag2017value}, aim to learn a linear decomposition of the joint Q-value, with each agent maintaining a network to approximate its own state-action values. This method ensures the sufficient condition for the IGM principle and becomes prevalent in MARL due to its simplicity and scalability, inspiring many subsequent approaches. Extending VDN, QMIX \cite{rashid2020monotonic} introduces a monotonic mixing network to enhance the expressiveness of the decomposed function class, ensuring that the optimal joint action maximizes the joint Q-value, aligning with the individual Q-values of each agent. However, QTRAN \cite{son2019qtran} aims to represent the entire IGM function class but faces computational intractability, requiring additional soft regularizers and not guaranteeing strict IGM consistency. QPLEX \cite{wang2020qplex} extends the IGM principle into the dueling network architecture, although it has potential limitations in scalability. These value decomposition methods illustrate the flexibility and adaptability of value-based CTDE approaches.

Although these algorithms have shown significant performance in many multi-agent cooperative tasks, their effectiveness relies heavily on the introduction of global state and the setup of centralized controllers. Unlike existing works, our algorithm through communication among agents, reliance on global state is eliminated, and the trainer is set within the agents themselves.

\subsection{MARL with Communication}

MARL with communication methods aim to achieve consensus and cooperation among multiple agents through learning effective communication strategies. Methods incorporating communication in MARL have significantly evolved from early approaches that relied on fixed, static broadcast mechanisms. Initial methods like reinforced inter-agent learning and differentiable inter-agent learning (RIAL \& DIAL) \cite{foerster2016learning} established foundational concepts in learning communication protocols among agents. Communication neural net (CommNet) \cite{sukhbaatar2016learning} further advanced communication learning by enabling agents to learn to broadcast messages. To alleviate the local policy burden caused by message flooding, methods like individualized controlled continuous communication model (IC3Net) \cite{singh2018learning} and attentional communication model (ATOC) \cite{jiang2018learning} introduced gating mechanisms to selectively target communication recipients. Despite these advancements, these mechanisms struggled with modeling directed communication, leading researchers to explore directed graph structures to specify communication targets more effectively. Approaches like graph convolutional reinforcement learning (DGN) \cite{jiang2018graph} and game abstraction mechanism based on two-stage attention network (G2ANet) \cite{liu2020multi} have shown promise in enhancing cooperation in dynamic and large-scale environments, respectively. However, none of these methods can precisely model the content of agent communication, as they primarily focus on selecting communication targets without considering the specific information being communicated. 

Recent research has made some progress in improving communication efficiency and mitigating the difficulty of policy learning. The first category focuses on generating meaningful messages for the message senders. A straightforward approach in this category is to treat raw local observations or the local information history as messages. For instance, Targeted multi-agent communication (TarMAC) \cite{das2019tarmac} achieves targeted communication through a soft-attention mechanism, where the sender broadcasts a key encoding the agents' properties, and the receiver processes all received messages for a weighted sum to make decisions. Nearly decomposable Q-functions (NDQ) \cite{wang2019learning} aims to generate minimal messages for different teammates, allowing them to learn decomposable value functions. NDQ optimizes the message generator using two information-theoretic regularizers to ensure expressive communication. The second category of work focuses on efficiently extracting the most useful messages at the receiver's end. An example is multi-agent communication via self-supervised information aggregation (MASIA) \cite{guan2022efficient}, which explicitly addresses the optimization of multiple received messages by introducing two self-supervised representation objectives. Multi-agent communication mechanism with Graph Information bottleneck optimization (MAGI) \cite{ding2023robust} introduces a robust communication learning mechanism, using graph information bottleneck optimization and information-theoretic regularizers to enhance the robustness and efficiency of multi-agent communication and coordination. These objectives aim to ensure that the received information representation abstracts the true states and predicts multi-step future information. These methods highlight the need for balancing the generation and reception of meaningful messages to facilitate effective communication and coordination among agents.

Despite significant advancements, a common limitation in current MARL communication methods is their reliance on sparse reinforcement learning rewards for training both communication and policy networks. This joint training approach often complicates local policies by introducing raw communication data, which increases the complexity of the learning process. Unlike existing methods, our approach focuses on information-level modeling, facilitating effective content extraction and selective integration of agents' local observations through communication for joint policy updates. Additionally, the use of pluggable modules improves computational efficiency, reduces policy network complexity, and enhances algorithm flexibility.

\subsection{Generative Adversarial Network}

Generative adversarial network (GAN) technology has significantly impacted artificial intelligence, especially in computer vision, by enabling the creation of highly realistic synthetic data. Introduced in 2014, GAN \cite{goodfellow2014generative} consist of a generator and a discriminator network that compete to improve their functions, resulting in data that closely mimics real-world samples. This innovative framework has led to significant progress in applications like image generation, enhancement, and inpainting. One of the earliest and most impactful improvements was the development of deep convolutional GAN (DCGAN) \cite{radford2015unsupervised}, which replaced fully connected layers with convolutional layers. This change enhanced both the stability and quality of generated images, setting a new standard for GAN architectures. Building on this, the introduction of super-resolution GAN (SRGAN) \cite{ledig2017photo} utilized a perceptual loss function based on high-level feature maps from pretrained networks. SRGAN enabled the generation of photo-realistic high-resolution images from low-resolution inputs, significantly improving visual quality over traditional methods. Further advancements are made with progressive growing of GAN (ProGAN) \cite{karras2017progressive}, a technique that gradually increased image resolution during training. Starting with low-resolution images and adding layers progressively led to more stable training, addressing issues like mode collapse and training instability. Additionally, the proposal of wasserstein GAN (WGAN) \cite{adler2018banach} introduces the wasserstein distance as a loss function, providing more meaningful gradients and improving training stability. Recent innovations have also tackled high-resolution image inpainting \cite{yang2017high}. A multi-scale neural patch synthesis approach combined deep convolutional networks for structural prediction with patch-based synthesis for texture generation, achieving coherent and sharp inpainting results, especially for high-resolution images. Additionally, the introduction of CycleGAN \cite{zhu2017unpaired} enables unpaired image-to-image translation by enforcing cycle consistency, broadening GAN's applicability in tasks like style transfer and domain adaptation. 

Overall, GAN has significantly advanced computer vision, particularly in generating high-fidelity visual content and enhancing image quality. However, the application of GAN in reinforcement learning remains underexplored. In our research, we find that in a MARL setting based on communication, GAN can effectively complement information. This means that through communication, agents can use GAN to combine received information with their observations, achieving global state information completion and addressing the issue of partial observability.

\section{Preliminaries}
\label{S_3}
\subsection{Decentralized Partially Observable Markov Decision Processes}

This paper models cooperative multi-agent systems with communication as decentralized partially observable markov decision processes (Dec-POMDP), which imposes partially observable settings on multi-agent markov decision processes. A Dec-POMDP can be defined by a tuple $\langle \mathcal{N}, \mathcal{S}, \mathcal{A}, \mathcal{P}, \Omega, \mathcal{O}, \mathcal{R}, \gamma, \mathcal{M} \rangle$, where $\mathcal N = \{1, \cdots ,n\}$ is the set of $n$ agents, $\mathcal{S}$ is the set of global states, $\mathcal{A}$ is the set of actions, $\mathcal{P}: \mathcal{S} \times \mathcal{A} \rightarrow \Delta(\mathcal{S})$ is the state transition probability distribution and $\Delta(\mathcal{S})$ is the set of probability distributions over $\mathcal{S}$, $\Omega$ is the set of observations, $\mathcal{O}: \mathcal{S} \times \mathcal{N} \rightarrow \Omega$ is the observation function that maps global states and agents to observations, $\mathcal{R}: \mathcal{S} \times \mathcal{A} \rightarrow \mathbb{R}$ is the reward function, $\gamma \in [0, 1)$ is the discount factor, $\mathcal{M}$ is the set of messages that agents can communicate. At each time step $t$, each agent $i \in \mathcal{N}$ receives an observation $o_t^i \in \Omega$ from the observation function $\mathcal{O}(s_t, i)$, where $s_t \in \mathcal{S}$ is the current global states. Each agent follows an individual policy $\pi^i(a_t^i | \tau_t^i, m_t^i)$, where $\tau_t^i = (o_1^i, a_1^i, \ldots, o_{t-1}^i, a_{t-1}^i, o_t^i)$ is the action-observation history of agent $i$, and $m_t^i \in \mathcal{M}$ is the message received by agent $i$ at time $t$. The joint action $\boldsymbol{a}_t = \langle a_t^1, \ldots, a_t^n \rangle$ leads to the next state $s_{t+1} \sim \mathcal{P}(s_{t+1} | s_t, \boldsymbol{a}_t)$, and the team receives a global reward $r_t = \mathcal{R}(s_t, \boldsymbol{a}_t)$. The objective is to find a joint policy $\boldsymbol{\pi} = \langle \pi^1, \ldots, \pi^n \rangle$ that maximizes the expected discounted return $ \mathbb{E}_{\boldsymbol{\pi}} \left[ \sum_{t=0}^\infty \gamma^t r_t \right]$. We use action-value functions to update policies, where in the distributed execution phase, each agent learns a $Q$ network $Q^i(\tau,a,m,s)$ \cite{mnih2015human} to approximate the action-value function $Q^i(s, a)$, and in the centralized training phase, the group of agents learns a mix network $Q_{\mathrm{tot}}(\boldsymbol{\tau},\boldsymbol{a},\boldsymbol{m},s;\boldsymbol{\theta})$ to approximate the global action-value function $Q_{\mathrm{tot}}(s,\boldsymbol{a})$. The parameters $\boldsymbol{\theta}$ are learned by minimizing the expected temporal difference (TD) error:

\begin{equation}
\label{e_1}
    \mathcal{L}(\theta)=\sum_{i=1}^b\left[\left(y_i^{\mathrm{tot}}-Q_{\mathrm{tot}}(\boldsymbol{\tau}_t,\boldsymbol{a}_t,\boldsymbol{m}_t,s_t;\boldsymbol{\theta})\right)^2\right],
\end{equation} where $y^{\mathrm{tot}}=r+\gamma\max_{\boldsymbol{a}_{t+1}}Q_{\mathrm{tot}}(\boldsymbol{\tau}_{t+1},\boldsymbol{a}_{t+1},\boldsymbol{m}_{t+1},s_{t+1};\boldsymbol{\theta}^{-})$, $\boldsymbol{\theta}^{-}$ are the parameters of a target network, and $b$ is the batch size of the transitions sampled from replay buffer $\mathcal{D}$.

\subsection{GAN in MARL with Communication}

GAN is a novel method of generative modeling that leverages discriminative models to generate realistic synthetic data. The GAN architecture consists of a generator network $G$ and a discriminator network $D$. We deploy the training process of the GAN model in a centralized training, where the generator obtains $b$ batches of local observations ${\boldsymbol{o}_t} = {\langle o_t^{1}, \ldots, o_t^{n} \rangle}$ as input from replay buffer $\mathcal{D}$. The discriminator is a binary classifier that outputs a scalar prediction representing the probability that input global states being real or fake. The discriminator is trained to minimize the cross-entropy loss for correctly classifying real and fake global states. On the other hand, the generator is trained to maximize the cross-entropy loss for the discriminator when classifying its generated global states as fake. The generator network $G$ and the discriminator network $D$ engage in a minimax game with the overall objective function:

\begin{equation}
\label{e_2}
    \min_D\max_G \left\{ -\mathbb{E}_{s\sim\mathcal{S}}\log D(\boldsymbol{s}_t)-\mathbb{E}_{o\sim\mathcal{O}}\log(1-D(G(\boldsymbol{o}_t)))\right\}.
\end{equation}

\section{Method}
\label{S_4}

\begin{figure*}
  \centering
  \centerline{\includegraphics[width=\textwidth]{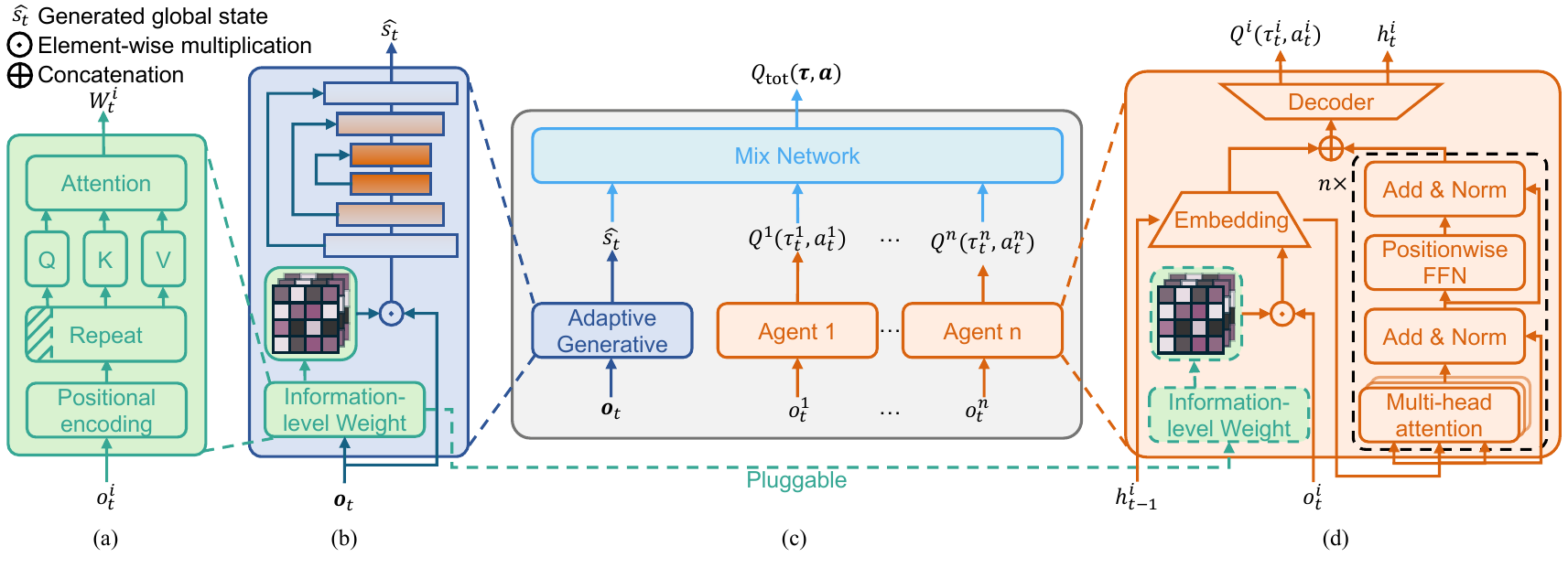}}
  \caption{The structure of PAGNet is differentiated by various colors representing different networks: green for the information-level weight network, purple for the adaptive generative network, blue for the mix network, and red for the Transformer-based decoder. (a) Information-level weight network. (b) Adaptive generative network. (c) The overall architecture. (d) Transformer-based decoder.}
  \label{f_1}
\end{figure*}

In this section, we explain the detailed structure and design of PAGNet, a MARL method with communication. The entire framework is depicted in Fig. \ref{f_1}. These innovations enhance cooperation performance while improving the efficiency of MARL. Information-level communication modeling addresses the issue of redundant information generated during agent communication and determines the specific information that needs to be transmitted. The adaptive generative networks address the problem of partial observability in MARL by communicating local observations through the information-level communication model to create global states using a generative model. The flexible pluggable modules integrate the information-level communication model with the generative model, allowing for switching between online and offline learning modes. The Transformer-based decoder addresses the challenge of integrating and extracting features from transmitted information post-communication.

PAGNet is an advanced framework that integrates value-based MARL into the CTDE paradigm. To enhance the capabilities of PAGNet, we have developed two additional networks: the information-level weight network for communication modeling and the adaptive generative network for global states generation. Moreover, we replace the multilayer perceptron (MLP) decoder in the policy network with a more efficient transformer architecture. The information-level weight network and the transformer-based decoder form the $Q$ network, which outputs $Q^i$ for each agent. The information-level weight network, the adaptive generative network, and the mix network collectively constitute the Network used to output $Q_{\mathrm{tot}}$. The information-level weight network not only facilitates information-level communication modeling but also serves as a pluggable module bridges centralized training and decentralized execution within PAGNet. We design a unique loss function to train information-level weight network with the adaptive generative network during centralized training.

\subsection{Communication Modeling with Information-level}
\label{S_4_1}

\paragraph{Communication Modeling} 

Effective communication among agents necessitates addressing three key aspects: timing, target, and content. Therefore, constructing a communication model at the information level is crucial for meeting these requirements and enabling agents to extract crucial information from redundant data. In this paper, we redefine the communication model by considering the local observations of each agent as communication information. These observations, after characteristic extraction, can be represented as a one-dimensional vector of length $l$, denoted as $m_t^i = o_t^i \in \mathbb{R}^l$. In the scenario of full communication, the complete information acquired by each agent at time $t$ is defined as $M_t^{i, -i} = \{o_t^i, m_t^{-i}\} \in \mathbb{R}^{n\times l}$, where the symbol $-$ denotes all other agents except agent $i$. Our objective is to learn a weight $W_t^{i, -i} \in [0,1]^{n \times l}$, which can simultaneously specify the timing, target, and content of communication. We redefine the weighted information $\boldsymbol{x}_t$ by referencing information-theoretic \cite{shwartz2024information}, as defined below:

\begin{equation}
\label{e_3}
x_t^i = \left(1 - W_t^{i, -i}\right) \cdot M_t^{i, -i} + W_t^{i, -i} \cdot \epsilon_t, \quad \epsilon_t \sim \mathcal{N}(0, I).
\end{equation}


\paragraph{Information-level Weight Network Architecture} 

\begin{figure*}
  \centering
  \centerline{\includegraphics[width=\textwidth]{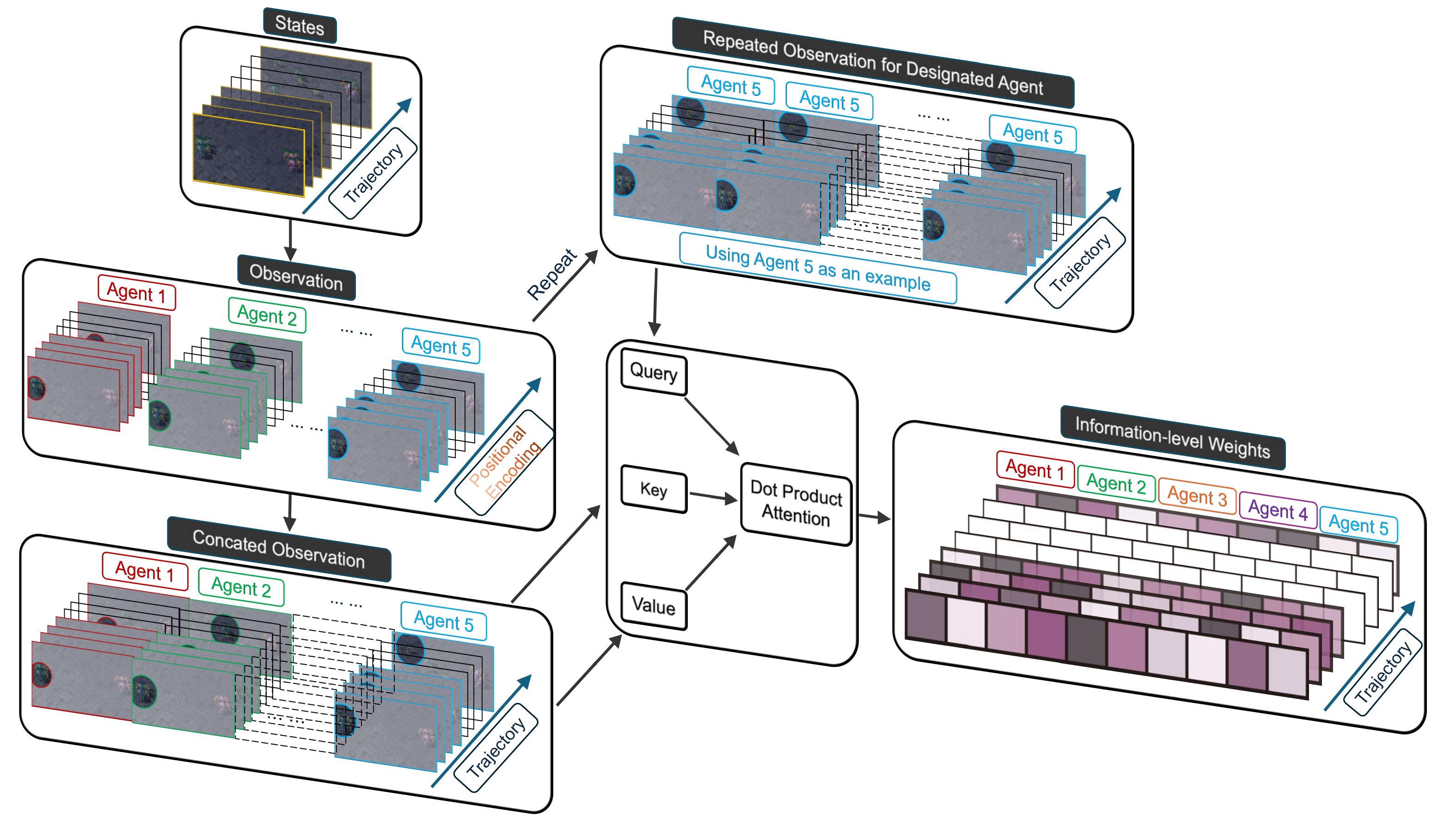}}
  \caption{Structure of information-level weight network. To provide a clearer demonstration, we set the number of agents to five and use the weight $W_t^5$ of Agent 5 as an example.}
  \label{f_A_2}
\end{figure*}

We design an information-level weight network to implement communication modeling. The structure of the information-level weight network is shown in the Fig. \ref{f_A_2}. The key feature of this network is to enable bidirectional communication at the information level, where $W_t^{i, j} \neq W_t^{j, i}$. The input to the network are the local observations of agents $\boldsymbol{o}_t$, and the output is the information-level weight $\boldsymbol{W}_t$. We use the local observations of agent $o_t^i$ as the query, and the information from agent $M_t^{i, -i}$ as the key and value. It should be noted that the agent's observations need to be repeated $n$ times to match the dimensions of the key. The dot product attention mechanism is applied to compute the information-level weight. This design aims to clarify the importance of the agent's local observations $o_t^i$ within the information received from other agents $M_t^{i, -i}$. To ensure that the learned information-level weight retains temporal information, positional encoding is added to the input representation before applying the attention mechanism. This encoding injects relative positional information from the MARL trajectory. The relevant calculation formulas are as follows:

\begin{equation}
\label{e_4}
\begin{aligned}
    \mathbf{Q}_t^{i} & = \mathrm{MLP}_{Q,K,V}([\underbrace{o_t^i + p^i, \ldots, o_t^i + p^i}_{n}]), \\
    \mathbf{K}_t^{i}, \mathbf{V}_t^{i} & = \mathrm{MLP}_{Q,K,V}\left([M_t^{i, -i} + \mathbf{p}]\right),
\end{aligned}
\end{equation} where query $\mathbf{Q}$, key $\mathbf{K}$, and value $\mathbf{V}$ are the concepts defined in the attention mechanism \cite{vaswani2017attention}. The positional encoding uses a matrix $\mathbf{P}\in\mathbb{R}^{n\times d}$ of the same shape, where the element in the $i^\text{th}$ row and $(2j)^\text{th}$ or $(2j+1)^\text{th}$ column is defined as $p^{i,2j} =\sin\left(\frac i{10000^{2j/d}}\right)$ and $p^{i,2j+1} =\cos\left(\frac i{10000^{2j/d}}\right)$. The information-level weight is computed using scaled dot product attention, as shown in the following: 

\begin{equation}
\label{e_5}    W_t^i=\sigma\left(\mathrm{MLP}_{Q,K,V}\left(\mathrm{softmax}\left(\frac{\mathbf{Q}_t^{i} {\mathbf{K}_t^{i}}^{\top}}{\sqrt{d}}\right)\mathbf{V}_t^{i}\right)\right),
\end{equation} where $d$ is the dimension of the query and key vectors, and $\sigma$ is activation function.

\subsection{Adaptive Generative Network Architecture}
\label{S_4_2}

\begin{figure*}
  \centering
  \centerline{\includegraphics[width=\textwidth]{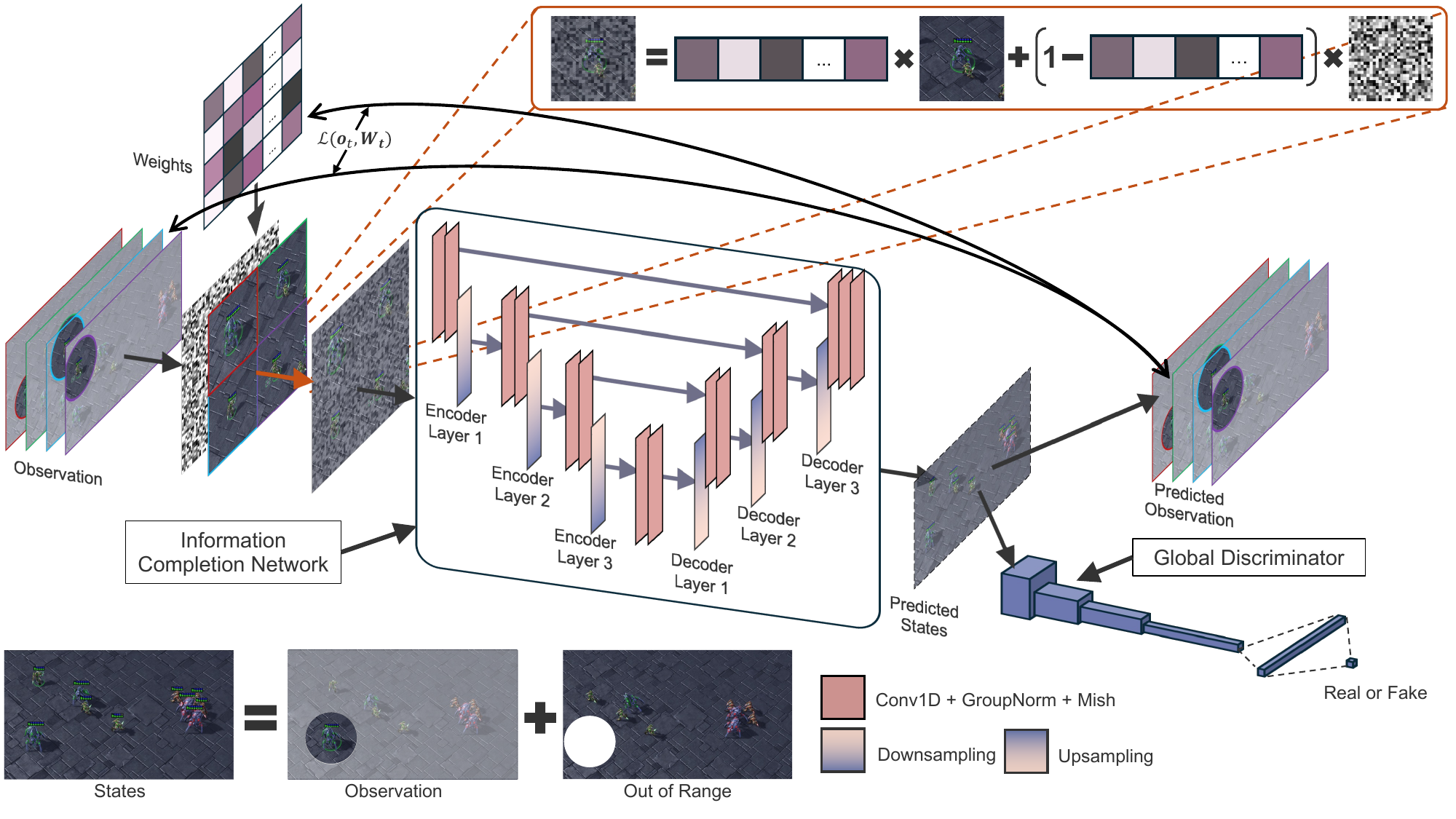}}
  \caption{Structure of adaptive generative network.}
  \label{f_2}
\end{figure*}

The purpose of designing the adaptive generation network is to achieve information completion by inputting agents' local observations (with $n$ input channels) and outputting global states (with 1 output channel). To ensure scalability, we integrate local observations of agents as a form of communication process. The information-level weight network introduced in Sec. \ref{S_4_1} does not rely on sparse reinforcement learning rewards for training. Instead, it assists the adaptive generation network in filtering out redundant local information, and it is jointly trained with the latter. Therefore, we set the content of communication to be local observations in this work. The adaptive generation network consists of two components: the information completion network and the global discriminator. The structure of the adaptive generative network is shown in Fig. \ref{f_2}. In the information completion network, the information-level weight network is incorporated because much of the information observed by agents is redundant. This redundancy becomes particularly significant when the number of agents increases. The use of the information-level Weight Network can alleviate the burden of information extraction in the information completion network.

\paragraph{Information Completion Network} 

In detail, we adopt the U-Net\cite{ronneberger2015u} as the base structure for modeling information completion. It consists of repeated one-dimensional convolutional residual blocks follows an encoder-decoder structure. The general architecture allows reducing memory usage and computational time by downsampling before further processing the information. Afterwards, the output is restored to the length of the global states by upsampling. The input of the completion network is the weighted information $\boldsymbol{x}_t$, which is calculated according to Eq. \eqref{e_3}. The output can be mathematically defined as $\widehat{s}_{t} = G(\boldsymbol{o}_t, \boldsymbol{W}_t)$, where $\widehat{s}_{t}$ represents the generated global states. An overview of the network architecture can be seen in Table \ref{t_1}.

\begin{table}
\caption{Information completion network architecture detailing each layer type, kernel size, and stride. After each convolution layer, there is a Mish layer \cite{misra2019mish}.}
\label{t_1}
  \centering
    \begin{tabular}{cccc}
    \toprule
    Layer         & Type       & Kernel & Stride\\
    \midrule
    encoder       & Linear     & -      & -\\
    \midrule
    \multirow{3}{*}{down[0]}    & \multirow{3}{*}{Downsample} & 5      & 1\\
                  &    & 5      & 1\\
                  &    & 3      & 2\\
    \midrule
    \multirow{3}{*}{down[1]}    & \multirow{3}{*}{Downsample} & 5      & 1\\
         &   & 5      & 1\\
         &   & 3      & 2\\
    \midrule
    \multirow{3}{*}{down[2]}    & \multirow{3}{*}{Downsample + Identity} & 5      & 1\\
         &   & 5      & 1\\
         &    & -      & -\\
    \midrule
    \multirow{2}{*}{mid}        & \multirow{2}{*}{Bottleneck} & 5      & 1\\
             &   & 5      & 1\\
    \midrule
    \multirow{3}{*}{up[0]}      & \multirow{3}{*}{Upsample}   & 5      & 1\\
          &     & 5      & 1\\
          &     & 4      & 2\\
    \midrule
    \multirow{3}{*}{up[1]}      & \multirow{3}{*}{Upsample}   & 5      & 1\\
           &     & 5      & 1\\
           &     & 4      & 2\\
    \midrule
    \multirow{3}{*}{up[2]}      & \multirow{3}{*}{Upsample + Identity}   & 5      & 1\\
           &     & 5      & 1\\
           &     & -      & -\\
    \midrule
    \multirow{3}{*}{decoder} & \multirow{3}{*}{Convolution + Linear}& 5      & 1\\
      &  & 1      & -\\
            &      & -      & -\\
    \bottomrule
  \end{tabular}
\end{table}

\paragraph{Global Discriminator Network} 

The global discriminator network has the objective of discerning whether the global states are real or generated through local observations. The networks are based on 1D convolutional neural networks that compress the global states into compact feature vectors. The outputs of the networks are fused through a concatenation layer that uses a sigmoid activation function to predict a continuous value between 0 and 1, representing the probability that the global state corresponds to the real situation rather than resulting from information completion.  An overview of the network architecture  can be seen in Table \ref{t_2}.

\begin{table}
\caption{Global discriminator network architecture detailing each layer type, kernel size, and stride.}
\centering
\label{t_2}
\begin{tabular}{cccc}
\toprule
Layer & Type & Kernel & Stride \\
\midrule
encoder & Linear & - & - \\
\midrule
\multirow{4}{*}{conv} & \multirow{4}{*}{Convolution} & 5 & 2 \\
  &   & 5 & 2 \\
  &   & 5 & 2 \\
  &   & 5 & 2 \\
\midrule
\multirow{2}{*}{decoder} & Flatten & - & - \\
  & Linear & - & - \\
\bottomrule
\end{tabular}
\end{table}

\paragraph{Training and Loss} 

To simultaneously train the adaptive generative network and the information-level weight network to realistically complete the global states, this training process employs two loss functions: a weighted mean squared error (MSE) loss for stability, and a GAN loss to enhance the realism of the results. To ensure that the local observations within an agent's observation range in the global state remain unchanged from the real local observations, the MSE loss is defined as:

\begin{equation}
\label{e_6}
    \mathcal{L}(\boldsymbol{o}_t, \boldsymbol{W}_t)=\parallel \mathcal{O}\left(G(\boldsymbol{o}_t, \boldsymbol{W}_t)\right) - \boldsymbol{o}_t\parallel^2,
\end{equation} where $\left\|\cdot\right\|$ is the Euclidean norm. The GAN loss transforms the neural network optimization into a min-max problem, where the discriminator network and the generator network are simultaneously updated at each iteration. Compared with Eq. \eqref{e_2}, the optimization objective can be formulated as follows:

\begin{equation}
\label{e_7}
\begin{split}
    \min_D\max_G & \left\{-\mathbb{E}_{\widehat{s}\sim\mathcal{S}}\log D(\widehat{\boldsymbol{s}_t}) \right.\\
    & \left. -\mathbb{E}_{o\sim\mathcal{O}}\log(1-D(G(\boldsymbol{o}_t, \boldsymbol{W}_t)))\right\}.
\end{split}
\end{equation}

By combining the two loss functions, the optimization becomes:

\begin{equation}
\label{e_8}
\begin{split}
    \min_D\max_G \left\{ - \mathbb{E}_{o\sim\mathcal{O}}\mathcal{L}(\boldsymbol{o}_t, \boldsymbol{W}_t) 
    - \mathbb{E}_{\widehat{s}\sim\mathcal{S}}\log D(\widehat{\boldsymbol{s}_t}) \right. \\
    \left. - \alpha\mathbb{E}_{o\sim\mathcal{O}}\log(1-D(G(\boldsymbol{o}_t, \boldsymbol{W}_t)))\right\},
\end{split}
\end{equation} where $\alpha$ is a weighing hyperparameter. 

\subsection{Pluggable Modules}
\label{S_4_3}

We propose an information-level weight network as a pluggable module in this paper. Since its parameters are updated according to Eq. \eqref{e_6}, pretraining of the weight network can be achieved if an offline dataset for MARL environments is available, allowing for a pluggable effect. The information-level weight network is present in both the distributed execution and centralized training processes. During distributed execution, only forward propagation is performed, regardless of whether pretraining is employed. However, during the centralized training process, if a pretraining method is not employed, it jointly updates its parameters with the adaptive generative network. Under the pluggable module setting, the expected TD error is modified from Eq. \eqref{e_1} to:

\begin{equation}
\label{e_9}
\begin{split}
    \mathcal{L}(\theta)=\sum_{i=1}^b\left[\left(y_i^{\mathrm{tot}}-Q_{\mathrm{tot}}(\boldsymbol{\tau}_t,\boldsymbol{a}_t,\boldsymbol{m}_t, \boldsymbol{W}_t, G(\boldsymbol{o}_t, \boldsymbol{W}_t);\boldsymbol{\theta})\right)^2\right],
\end{split}
\end{equation} where $y^{\mathrm{tot}}=r+\gamma\max_{\boldsymbol{a}_{t+1}}Q_{\mathrm{tot}}$.

\subsection{Transformer-based Decoder}

We design a Transformer-based decoder to replace the MLP network architecture in the distributed execution process. An overview of the network architecture can be seen in Fig. \ref{f_1}(d). After communicating through the information-level weight network, the agents obtain weighted information from other agents. Although this information undergoes filtering and extraction by the information-level weight network, when the number of agents is relatively large, the agents possess significantly more information compared to their local observations. Therefore, we select the Transformer architecture, which demonstrates outstanding performance in handling sequential problems, to enable feature extraction from the available information. This further assists the agents in making decisions.

Initially, the agents' local observations undergo the information-level weight network according to Eq. \eqref{e_3}, and the agents obtain the weighted information. Subsequently, the historical information is initialized, and the input is preprocessed by embedding the weighted information $\boldsymbol{x}_t$ and the historical information $h^i_{t-1}$. Next, the output of the embedding is used as input for the Transformer-based decoder. This output contains both the historical state information and the fused information, which will be fed to the Transformer block at the next time step. Each such Transformer block is composed of self-attention mechanism, positionwise feed-forward networks (FFN), residual connection and layer normalization to prevent gradient vanishing and network degradation as the depth increases. To achieve a better balance between performance and memory consumption, it is common to use two Transformer blocks. The output of Transformer blocks and the previous history information concatenated together go through a fully-connected layer to get the individual action-value function $Q^i(\tau_t^i,a_t^i)$ and the hidden state at the next moment $h_t^i$.

\subsection{The Overall Flow of PAGNet for Training}

To illustrate the process of training, the overall training flow of PAGNet is shown in Alg. \ref{alg_1}. Lines 5 to 15 describe the entire training process. Specifically, lines 6 to 9 detail the data collection in the replay buffer. Line 10 updates the adaptive generative network and the information completion network by calculating the loss using Eq. \eqref{e_8}. Line 11 updates the Transformer-based decoder and mix network based on the MARL rewards by calculating the loss using Eq. \eqref{e_9}.

\begin{algorithm}
   \caption{Overall Training Framework}
   \label{alg_1}
\begin{algorithmic}[1]
   \STATE Initialize replay buffer $\mathcal{D}$.
   \STATE Initialize $Q$ network with random parameters $\theta$ and latent model with same parameter.
   \STATE Initialize adaptive generative network with random parameters $\phi$ and information-level weight network with random parameters $\psi$.
   \STATE Load information-level weight model parameters if it exist.
   \FOR{$\text{episode}=1$ {\bfseries to} $M$}
   \STATE Roll out one trajectory $\tau$ with $\epsilon$-greedy policy in the environment.
   \STATE Store the trajectory $\tau$ in replay buffer $\mathcal{D}$.
   \IF{$\lvert \mathcal{D} \lvert$ is larger than batch size}
   \STATE Sample a minibatch $\mathcal{B}$ from  from $\mathcal{D}$.
   \STATE Compute adaptive generative network loss:\\
   \vspace{-1.5em}
    \begin{equation*}
    \begin{split}
    \min_D\max_G \left\{ - \mathbb{E}_{o\sim\mathcal{O}}\mathcal{L}(\boldsymbol{o}_t, \boldsymbol{W}_t) 
    - \mathbb{E}_{\widehat{s}\sim\mathcal{S}}\log D(\widehat{\boldsymbol{s}_t}) \right. \\
    \left. - \alpha\mathbb{E}_{o\sim\mathcal{O}}\log(1-D(G(\boldsymbol{o}_t, \boldsymbol{W}_t)))\right\}.
    \end{split}
    \end{equation*}
    \vspace{-1em}
   \STATE Compute the reinforcement learning loss:\\
   \vspace{-1.5em}
   \begin{equation*}
    \begin{split}
    \mathcal{L}(\theta) = \sum_{i=1}^b \Bigg[&\bigg(y_i^{\mathrm{tot}} -
    Q_{\mathrm{tot}}(\boldsymbol{\tau}_t, \boldsymbol{a}_t, \boldsymbol{m}_t, \\
    &\boldsymbol{W}_t, G(\boldsymbol{o}_t, \boldsymbol{W}_t); \boldsymbol{\theta})\bigg)^2\Bigg].
    \end{split}
    \end{equation*}
    \vspace{-1em}
   \STATE Update $\theta$, $\phi$ and $\psi$ and save the information-level weight model parameters $\psi$.
   \ENDIF
   \STATE Update target network parameters.
   \ENDFOR
\end{algorithmic}
\end{algorithm}

\section{Experiment}
\label{S_5}

We conduct experiments on various benchmarks with differing levels of communication requests, comparing PAGNet against several baselines, including both methods without communication and state-of-the-art communication techniques. QMIX \cite{rashid2020monotonic} is a strong baseline without communication that has demonstrated outstanding performance across various multi-agent benchmarks \cite{papoudakis2020benchmarking}. DGN \cite{jiang2018graph} leverages graph convolutional networks with relational kernels to capture dynamic agent interactions. G2ANet \cite{liu2020multi} constructs a sparse communication interaction graph using a two-stage attention mechanism. MASIA \cite{guan2022efficient}, proposed in 2022, features a permutation-invariant message encoder and extraction mechanism for the self-supervised aggregation of communication data. Additionally, we design two ablation experiments based on PAGNet: PAGNet with full communication (PAGNet\_FC) and PAGNet with pretrain (PAGNet\_PT). In PAGNet\_FC, the information-level weight network is replaced with a full communication structure. In PAGNet\_PT, we deploy the pretrained information-level weight network and adaptive generative network. All algorithms utilize the EPyMARL \cite{papoudakis2020benchmarking} framework implementation. We evaluate PAGNet with multiple state-of-the-art baselines on tasks including LBF \cite{papoudakis2020benchmarking}, Hallway \cite{wang2019learning}, and the the SMAC \cite{samvelyan2019starcraft}. Evaluation entails reporting mean performance alongside a 95\% confidence interval across 5 random seeds.


\subsection{Details about Benchmarks}

We evaluate PAGNet against multiple state-of-the-art baselines in three testing benchmarks. In this section, we describe the details of these benchmarks.

\paragraph{Level-based Foraging} 

\begin{figure*}
  \centering
  \centerline{\includegraphics[width=\textwidth]{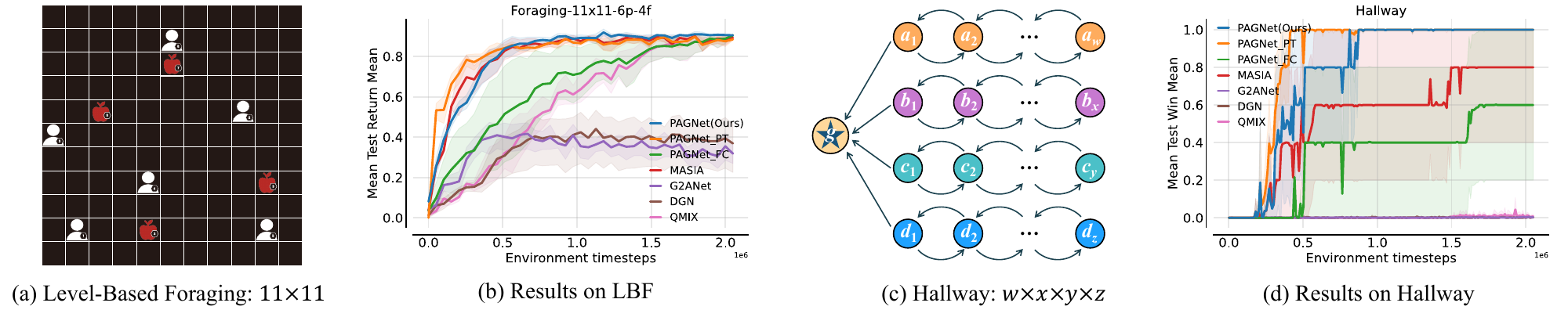}}
  \caption{(a) Illustrations of LBF tasks. (b) Average returns on LBF. (c) Illustrations of Hallway tasks. (d) Average battle win rate on Hallway.}
  \label{f_4}
\end{figure*}

LBF is a partially observable grid-world game where agents and food items are initialized with random skill levels. Each agent's action space includes movement in four directions, loading food, and a ``none'' action. A group of agents can collect food if they all choose the loading food action and their combined skill levels are greater than or equal to the food's level. Agents receive a reward proportional to the food's level. The goal is to maximize the global return within a limited horizon, with the maximum return normalized to one. Fig. \ref{f_4}(a) illustrates our task, where six agents cooperate to collect four food in an $11 \times 11$ grid world within a limited horizon. Agents have a restricted vision range of 2. The differing skill levels between agents and food necessitate strong coordination to achieve a high return.

\paragraph{Hallway}

Hallway (Fig. \ref{f_4}(b)) is a sparse reward cooperative environment with four agents randomly initialized at states $a_1$ to $a_w$, $b_1$ to $b_x$, $c_1$ to $c_y$, and $d_1$ to $d_z$. Each agent can only observe its own position and can select actions from moving left, moving right, and staying still. The episode ends if some agents arrive at state $g$, and they win and receive a reward of 10 only when reaching state $g$ simultaneously. The horizon is set to $\max(w, x, y, z) + 10$ to avoid infinite loops. We set $w = 4, x = 6, y = 8, z = 10$ to make simultaneous arrival difficult, as agents need strong coordination and communication to win in this partially observable task.

\paragraph{StarCraft Multi-agent Challenge}

SMAC consists of a set of StarCraft II micro scenarios designed to evaluate how well independent agents can learn coordination to solve complex tasks. These scenarios are complicated, in which the ally agents are necessitated to learn no less than one micromanagement technology to defeat the enemy agents. We describe in detail the scenarios with communication as introduced in NQD \cite{wang2019learning}. The remaining scenarios (easy, hard, and super hard) can be found in the SMAC paper \cite{samvelyan2019starcraft}.

1o2r\_vs\_4r: An Overseer detects four Reapers. To win, its two allied Roaches must find and eliminate the Reapers. At the start, the Overseer and Reapers spawn randomly, and the Roaches start at another random point. Since only the Overseer knows the enemies' locations, it must relay this information to the Roaches to win.

1o10b\_vs\_1r: On a cliff-filled map, an Overseer detects a Roach. To win, the Overseer's ten allied Banelings must eliminate the Roach. The Overseer, Roach, and Banelings all spawn at random locations. With minimal communication, the Overseer must signal its position to the Banelings. This task tests our method in complex scenarios.

\subsection{Overall Performance Comparisons}
\label{S_5_1}

\begin{figure*}
  \centering
  \centerline{\includegraphics[width=\textwidth]{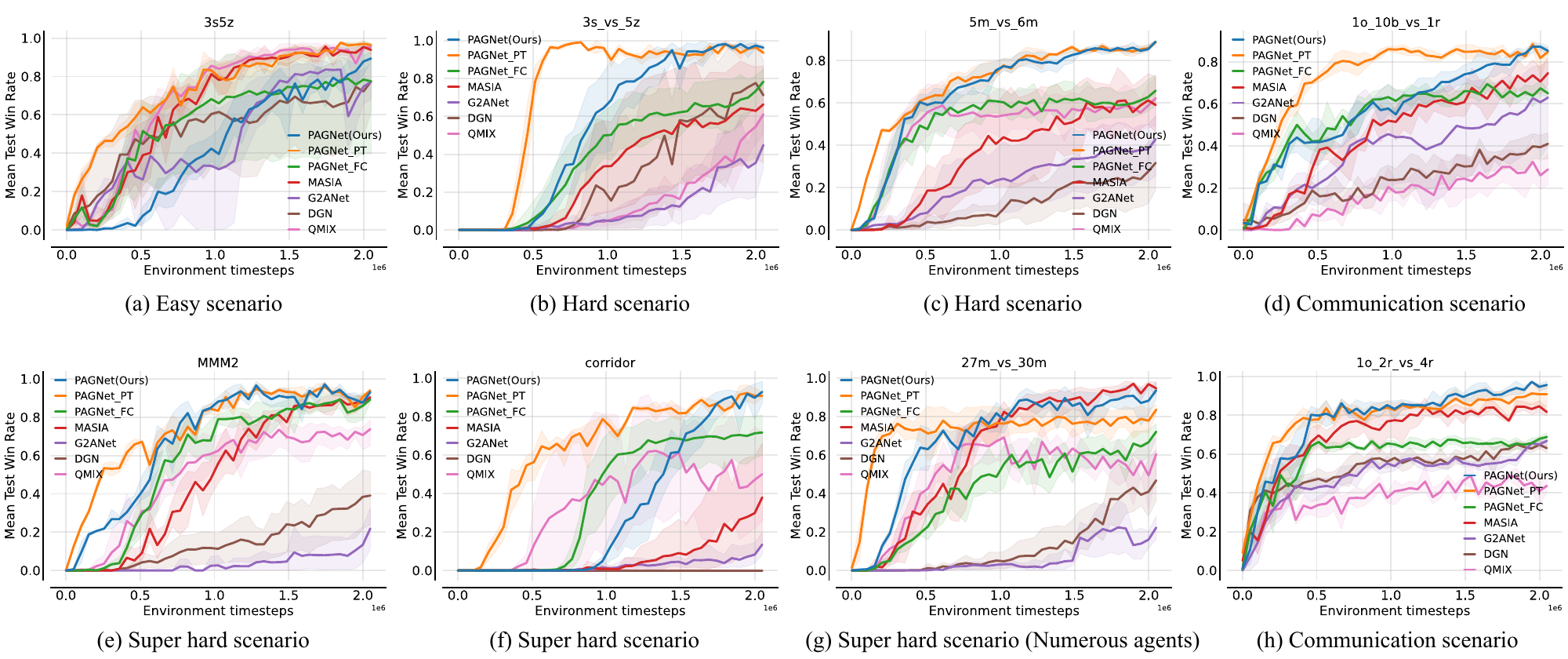}}
  \caption{Performance comparisons with baselines on SMAC.}
  \label{f_3}
\end{figure*}

To thoroughly assess and compare the overall performance of all algorithms, Fig. \ref{f_3}  shows the average test win rates across all eight maps in the SMAC scenario. Overall, PAGNet achieves one of the highest average test win rates across all scenarios upon convergence. Although the overall performance of MASIA is slightly inferior to that of PAGNet, its convergence speed and overall win rate upon convergence are greater than those of DGN and G2ANet due to communication information aggregation. Although DGN and G2ANet perform similarly to MASIA in scenarios that require communication, they converge to lower performance than QMIX in scenarios without communication requirements. This discrepancy can be attributed to their inaccurate communication modeling, resulting in ineffective management of redundant information and hampering the training process of the reinforcement learning policy networks. In the hard scenarios, PAGNet surpasses the other algorithms' average test win rate by approximately 20\%, likely benefiting from the superior ability of the Transformer-based decoder to represent policy networks compared to that of the MLP. In super hard scenarios with relatively larger numbers of agents, especially 27m\_vs\_30m, PAGNet maintains a higher convergence speed, thanks to the information-level weight network. QMIX performs the worst in scenarios requiring communication without a communication mechanism. while PAGNet demonstrates superior performance in MARL with communication, further validating the reasonableness of its network architecture. The results of the ablation experiments indicate that, compared to PAGNet, the pretrained PAGNet\_PT achieves the fastest convergence speed. Conversely, PAGNet\_FC, which lacks an information-level weight network, significantly reduces the convergence speed and performance due to its inability to extract important information from redundant data.

In the sparse-reward LBF scenario (Fig. \ref{f_4}(b)), MARL methods with communication such as G2ANet and DGN struggle when foods are sparsely distributed. In the Hallway scenario (Fig. \ref{f_4}(d)), which requires frequent communication, methods without communication such as QMIX fail entirely. Other communication-based MARL methods also perform poorly or fail in this environment, indicating that inappropriate message generation or selection can severely impair learning. Conversely, our algorithm excels in all three scenarios, demonstrating its superiority.

\subsection{Performance of Communication}
\label{S_5_2}

\begin{figure*}
  \centering
  \centerline{\includegraphics[width=\textwidth]{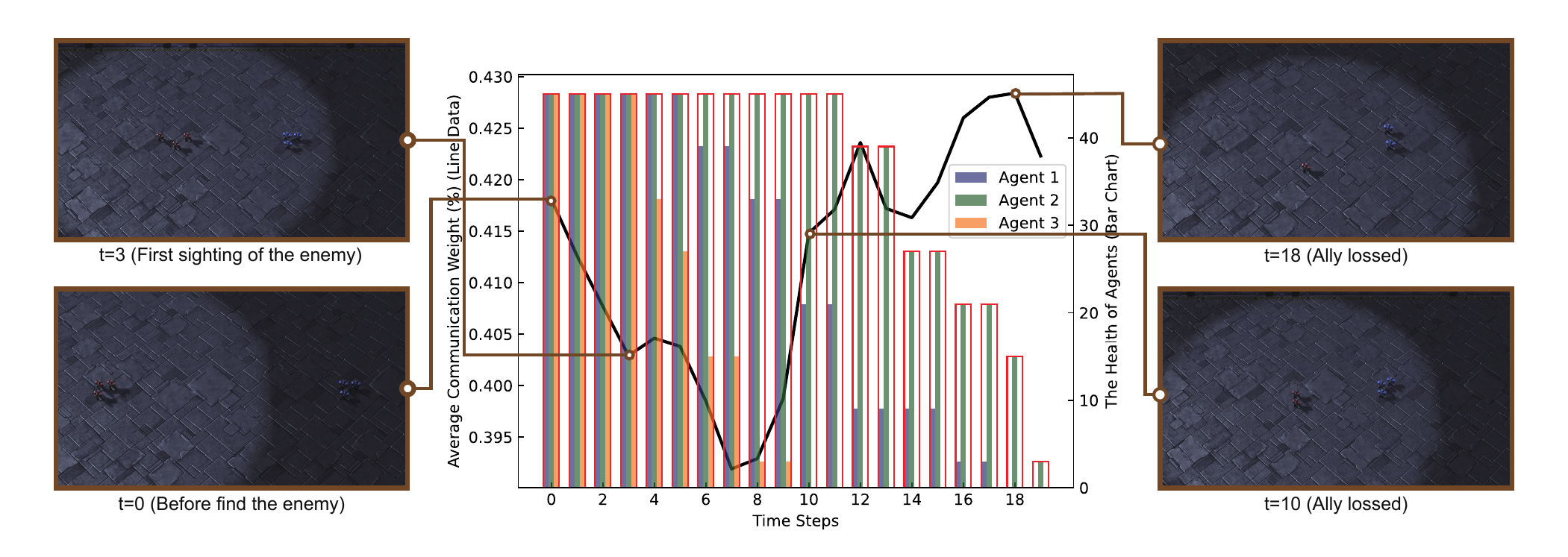}}
  \caption{Information-level weight visualization and analysis along the MARL trajectory. The line graph's y-axis (Average Communication Weight (\%)) is on the left side, while the bar graph's y-axis (The Health of Agents) is on the right side.}
  \label{f_5}
\end{figure*}

To examine the communication strategies acquired by the information-level weight network throughout the MARL trajectory, a visualization analysis is conducted on the network post-training. The average communication weights and health levels of the agents are plotted over the entire MARL trajectory, as depicted in Fig. \ref{f_5}. The x-axis represents the trajectory steps in time, while the left y-axis represents the average communication weight shown as a black line graph. The right y-axis represents the health levels of the agents represented as a bar graph. The line graph generally demonstrates a trend of initial decrease followed by an increase. Four specific points are selected for detailed analysis: $t=0, 3, 10, 18$. In addition, visualizations of the SMAC scenario at each of these time points are provided, with black areas representing regions outside the agents' field of view.

At $t=0$, the scenario has just commenced, and no enemies are visible to the allies. Communication is frequent during this stage, as agents need to exchange all available information regarding their allies. From $t=0$ to $t=3$, although enemies are still not visible, the communication frequency decreases due to the lack of new information to exchange, except for critical details such as the relative positions of agents. At $t=3$, an enemy comes into view for one of our agents, resulting in increased communication. By $t=10$, one of our agents has either been lost or has a health value of zero, leading to heightened communication among the remaining agents. At $t=18$, another agent is lost or has a health value of zero. With only one agent remaining, the communication weight begins to decrease. In summary, the visualization results indicate that the information-level weights learned by PAGNet exhibit a certain level of interpretability.

\subsection{Performance of the Information Completion}
\label{S_5_3}

\begin{figure*}
  \centering
  \centerline{\includegraphics[width=\textwidth]{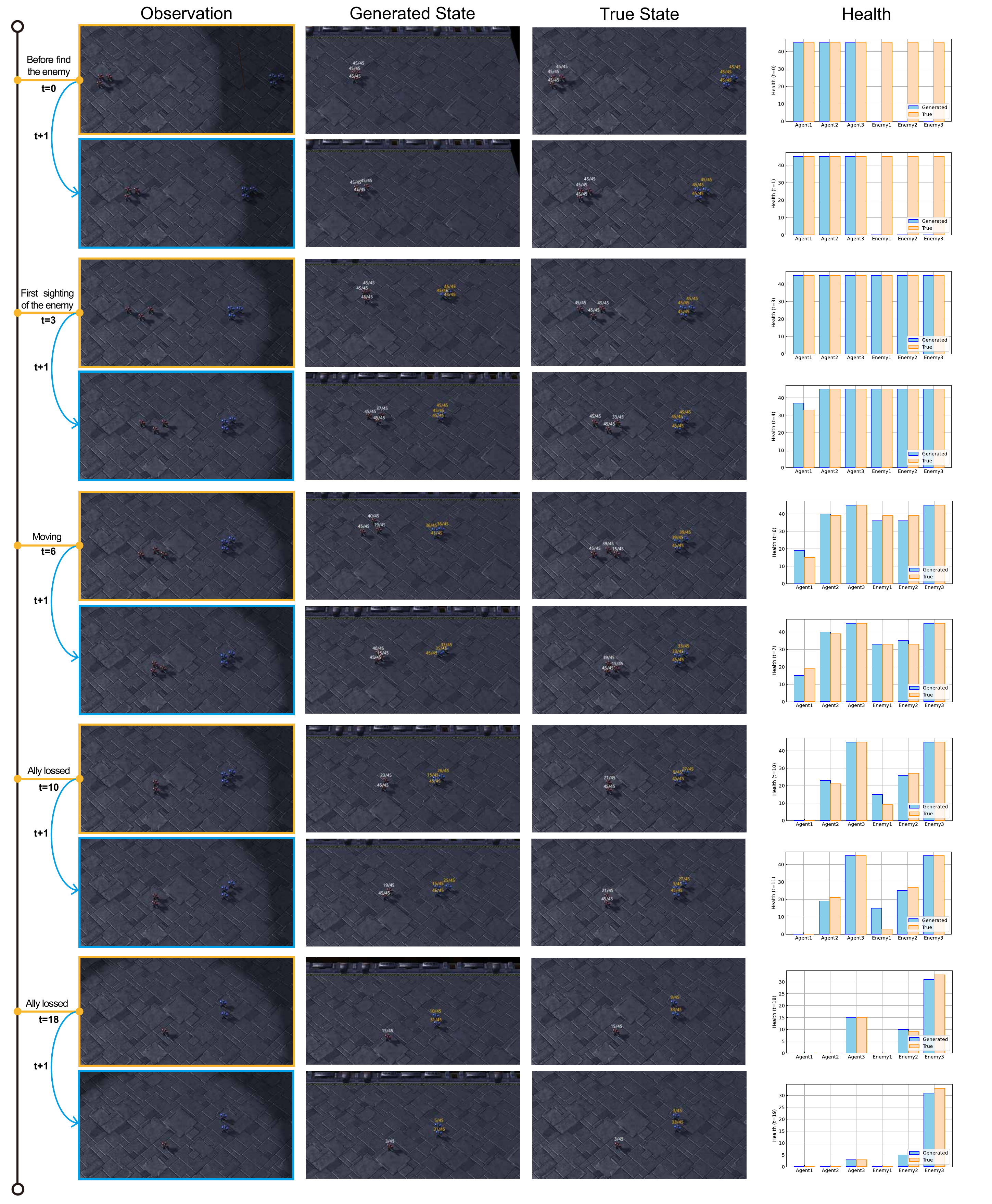}}
  \caption{Information completion visualization and analysis along the MARL trajectory. In the generated state and true state, the numbers above the agents indicate their health values. For example, 33/45 means the agent's current health is 33, and the initial health was 45.}
  \label{f_A_1}
\end{figure*}

To assess the impact of the adaptive generative network on agent decision-making in comparison to real global states, we visualize the agents' local observations, generated global states, real global states, and agents' health levels throughout the MARL trajectory. These visualizations can be found in the Fig \ref{f_A_1}. Overall, the generated global states exhibit temporal continuity in agents' relative positions and health statuses. At $t=0$, there are no enemies within the allies' view, resulting in a lack of enemy information in the generated state. However, at $t=3$, when enemies enter the agents' fields of view, the generated global state closely resembles the real global state. By $t=6$, there are changes in agent formation, but the relative positions in the generated global state remains consistent with those in the real global state. Additionally, when an agent is lost or its health reached zero, the generated global state also omits information about that particular agent. In summary, the generated global states closely matched the real states, effectively capturing information about agents' positions, health, formations, and enemy presence throughout the trajectory.

\subsection{Scalability of PAGNet}

\begin{figure}
  \centering
  \centerline{\includegraphics[width=2.5in]{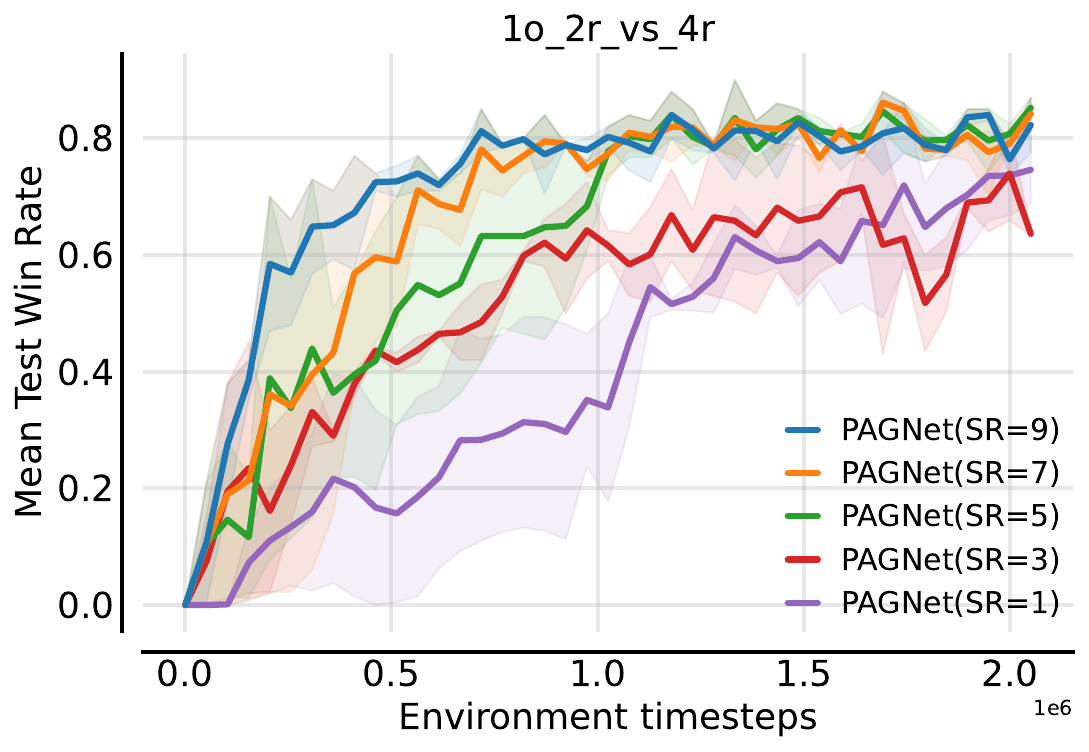}}
  \caption{Performance comparisons with varying sights (SR means the sightrange).}
  \label{f_6}
\end{figure}

To demonstrate the scalability of PAGNet, we conducte additional experiments from different sights range. First, in the 1o\_2r\_vs\_4r scenario, we gradually reduce the agents’ field of view from 9 to 1, creating tasks that require varying levels of communication. This allows us to assess PAGNet's ability to generalize for agents with limited observations. The training results are depicted in Fig. \ref{f_6}. As the agents’ visibility range diminishes, the convergence difficulty for PAGNet increases. However, the post-convergence performance remains consistent, indicating that the adaptive generative network successfully captures critical global state information even under restricted visibility conditions.

\subsection{Details of Hyperparameters}
\paragraph{Hyperparameters of PAGNet}

The selection of the additional parameters introduced in PAGNet is listed in Table \ref{t_3}. Our PAGNet implementation is built upon EPyMARL with SMAC 2.4.10 \cite{papoudakis2020benchmarking}, utilizing its default hyperparameter configurations. As an illustration, all buffer capacities are set to 5000, and the learning rate is fixed at 0.0005. We employ the default $\epsilon$-greedy action selection strategy for each method, where $\epsilon$ is linearly annealed from 1 to 0.05 over the initial 50,000 timesteps. We use this set of parameters in all experiments shown in this paper except for the ablations.

\begin{table}
\caption{The special hyperparameters of PAGNet.}
\label{t_3}
  \centering
    \begin{tabular}{lc}
    \toprule
    Hyperparameter & Value\\
    \midrule
    weight hidden dim    & 64\\
    weight dropout & 0.1\\
    decoder attention heads    & 4\\
    attention decoder dim     &  128\\
    number layers  & 2\\
    alpha      &  0.0004\\
    discriminator embedding dim  & 128\\
    \bottomrule
  \end{tabular}
\end{table}

\paragraph{Hyperparameters Tuning} 

For each benchmarks, we evaluate approximately the same number of hyperparameter combinations across all applicable algorithms to ensure consistency. The range of hyperparameters evaluated in each environment for the respective algorithms is presented in Table \ref{t_4}. To produce the results showcased in this work, we select the combination of hyperparameters that achieves the maximum average evaluation over three different seeds. While the hyperparameter search is limited in the SMAC environment to reduce computational cost, several of the evaluated algorithms have been previously assessed in SMAC, with their optimal hyperparameters publicly available in their respective papers. For one of its tasks in each environment, the parameters of each algorithm are optimized, and these optimized parameters are then kept constant for the remaining tasks within that same environment. Each combination of hyperparameters is evaluated across three different seeds.

\begin{table*}
\caption{Range of hyperparameters for PAGNet that is evaluated in each environment.}
\label{t_4}
  \centering
    \begin{tabular}{lccc}
    \toprule
    Hyperparameter & SMAC & LBF & Hallway\\
    \midrule
    weight hidden dim    & 64/128 &  32/64/128 & 32/64/128 \\
    weight dropout & 0/0.1 & 0/0.1/0.2 &  0/0.1/0.2\\
    decoder attention heads  & 2/4 & 2/4/6 & 2/4/6\\
    attention decoder dim  & 64/128 & 32/64/128 & 32/64/128\\
    number layers  & 1/2 & 1/2/4 & 1/2/4\\
    alpha & 0.0004/0.0005 & 0.0003/0.0004/0.0005 & 0.0003/0.0004/0.0005\\
    discriminator embedding dim  & 64/128 & 32/64/128 & 32/64/128\\
    \bottomrule
  \end{tabular}
\end{table*}

\subsection{Experimental Details}

The experimental evaluation is conducted on a desktop machine equipped with eight NVIDIA V100 GPUs. The total training time for our model varies depending on the complexity of the experimental environment. For all the performance curves present in our paper, we periodically pause the training process every $M$ timesteps and assess the performance over $N$ episodes, employing a decentralized greedy action selection strategy. Specifically, the $(M, N)$ pairs for LBF, Hallway, and SMAC are (50K, 100), (10K, 100), and (50K, 100), respectively. We evaluate the test win rate, which represents the percentage of episodes where the agents successfully complete the task within the prescribed time limit, across $N$ testing episodes for all tasks.

\subsection{Computational Cost of All Algorithms}

\begin{figure}
  \centering
  \centerline{\includegraphics[width=2.5in]{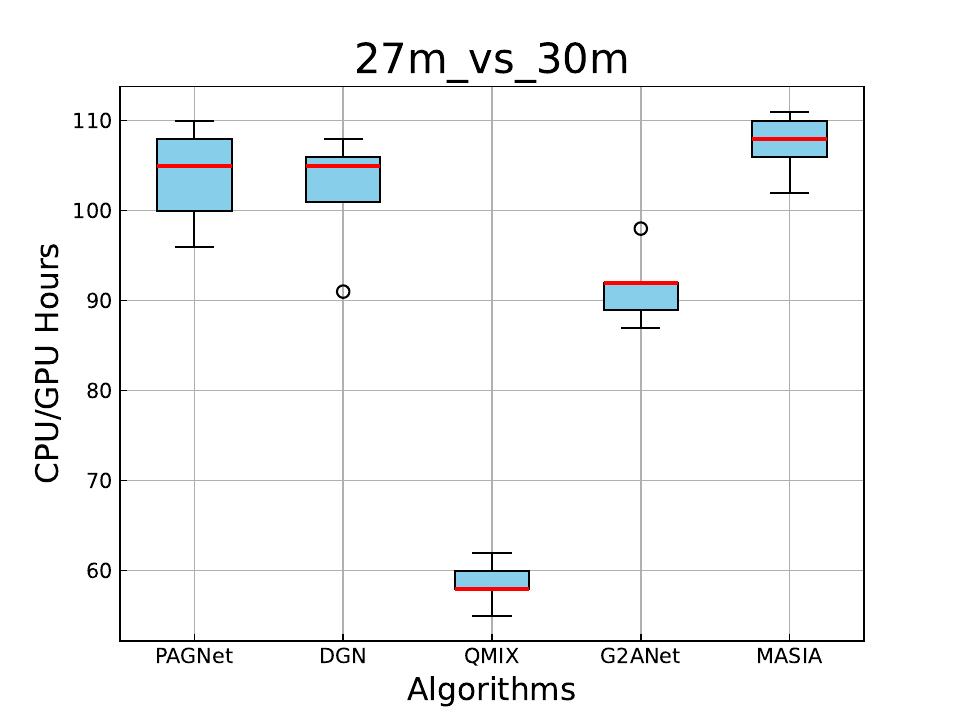}}
  \caption{Computational cost of all algorithms in 27m\_vs\_30m scenario.}
  \label{f_7}
\end{figure}

We conduct extensive evaluations on the computational requirements of PAGNet during the development. Our findings indicate that the additional computational overhead introduced is manageable and scales reasonably with the number of agents and problem complexity. The decentralized nature of our approach, where agents communicate and update their policies independently, mitigates the computational burden compared to centralized approaches. We test our model in the most agent-intensive scenario, 27m\_vs\_30m, with results shown in the Fig \ref{f_7}. The figure shows that the average computational cost of PAGNet is roughly the same as 
that of MASIA.

\section{Conclusion}
\label{S_6}

In this work, we propose PAGNet, a novel framework that incorporates generative models into multi-agent reinforcement learning with communication. Our work proposes several critical innovations to address the limitations of existing methods: 1) An information-level weight network for selective message extraction. 2) Second, we implement adaptive generative networks that complete global state representations based on weighted local observations. 3) Finally, we design a flexible structure that separates communication and policy training, resulting in improved computational efficiency and bridging the gap between centralized training and decentralized execution. The superior performance of PAGNet is demonstrated through extensive evaluations conducted on various cooperative MARL benchmarks. Additionally, a detailed analysis of the emergent communication patterns and generated global states offers valuable insights into the mechanisms and interpretability of PAGNet. We anticipate future work that establishes benchmarks for MARL with image-based inputs and addresses scalability challenges associated with environments containing hundreds or thousands of agents.

\bibliographystyle{IEEEtran}
\bibliography{ref}

\end{document}